# Influence of morphology on the plasmonic enhancement effect of Au@TiO$_2$ core-shell nanoparticles in dye-sensitized solar cells

*Wei-Liang Liu,[1] Fan-Cheng Lin,[1] Yu-Chen Yang,[2] Chen-Hsien Huang,[1] Shangjr Gwo,[2,3] Michael H. Huang[1] and Jer-Shing Huang[1,3,4*]*

1. Department of Chemistry, National Tsing Hua University, Hsinchu 30013, Taiwan
2. Department of Physics, National Tsing Hua University, Hsinchu 30013, Taiwan
3. Center for Nanotechnology, Materials Sciences, and Microsystems, National Tsing Hua University, Hsinchu 30013, Taiwan
4. Frontier Research Center on Fundamental and Applied Sciences of Matters, National Tsing Hua University, Hsinchu 30013, Taiwan

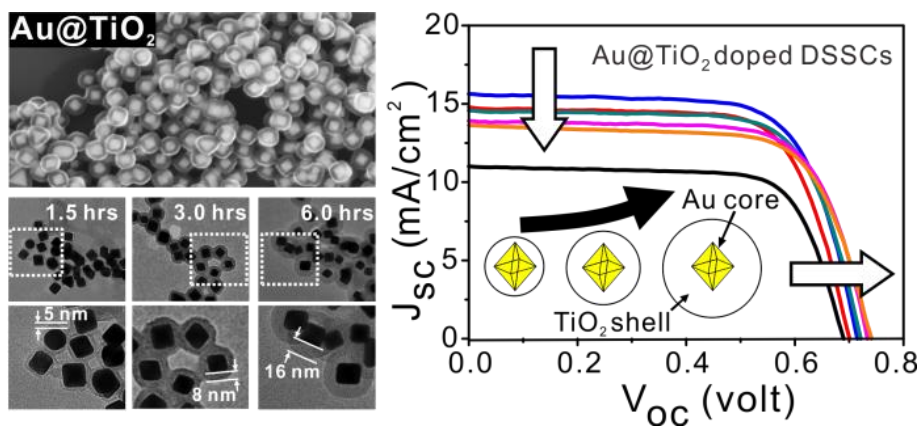

AUTHOR ADDRESS: jshaung@mx.nthu.edu.tw

KEYWORDS  Plasmonics, field enhancement, core-shell nanoparticles, dye-sensitized solar cells, metal-semiconductor composite, photovoltaics

ABSTRACT

Plasmonic core-shell nanoparticles (PCSNPs) can function as nanoantennas and improve the efficiency of dye-sensitized solar cells (DSSCs). To achieve maximum enhancement, the morphology of PCSNPs need to be optimized. Here we precisely control the morphology of Au@TiO$_2$ PCSNPs and systematically study its influence on the plasmonic enhancement effect. Enhancement mechanism was found to vary with the thickness of TiO$_2$ shell. PCSNPs with thinner shell enhance the current due to plasmonic effect, whereas particles with thicker shell improve the voltage due to increasing semiconducting character. Wavelength-independent enhancement in the visible range was observed and attributed to plasmonic heating effect. PCSNPs with 5-nm shell give highest efficiency enhancement of 23%. Our work provides a new synthesis route for well-controlled Au@TiO$_2$ core-shell nanoparticles and gains insight into the plasmonic enhancement in DSSCs.

In dye-sensitized solar cells (DSSCs), the photon-to-current conversion starts from the absorption of light by sensitizing dye molecules. The excited sensitizers then oxidize through ultrafast electron injection into the conduction band of adjacent semiconductor. Oxidized dyes are quickly reduced to its neutral form by the reducing agent in the electrolyte, e.g. iodide ions in $I^-/I_3^-{}_{(aq)}$ electrolytic system. The electrons travel through interfaces and external circuits and eventually recombine with the oxidant of the electrolyte at the counter electrode. Overall, the incident light creates a voltage determined by the potential difference between conduction band of the semiconductor and the reduction potential of the oxidant in the electrolyte. In a shorted circuit, such a potential can flow current and generate electrical power.[1, 2] Since multiple steps are involved, the performance of DSSCs can be enhanced by diversely different optimization approaches,[3] including enlarging dye absorption spectrum and suppressing recombination rate by molecular design,[4-6] decreasing the internal impedance of the cell by resistance engineering,[7-9] optimizing the electrolytic redox system,[10, 11] and enhancing the light harvesting efficiency by increasing scattering[12, 13] or by using plasmonic enhancement effect.[14-21]

At resonance, plasmonic nanoparticles exhibit localized surface plasmon resonance (LSPR) and function as optical nanoantennas that concentrate light to a sub-wavelength area.[22] Such a highly confined and enhanced field greatly improves the mismatch between photons and molecules, and thereby enhances the nanoscale light-matter interactions.[22-24] Since the intensity of enhanced optical near field decays rapidly from the metal-dielectric boundaries,[25] a shorter distance from the metal surface is generally preferred in order to gain stronger enhancement. However, if the distance is too short, the metallic surface may quench the excited molecules by introducing additional non-radiative decay channels, such as electron or energy transfer to the metal surface.[26-28] Surface plasmon resonance can also efficiently convert electromagnetic energy into heat, capable of increasing the local temperature and inducing nanoscale phase transition.[29-32] Recently, such plasmonic thermal effect has been used in thermal therapy, photon-induced drug release and solar vapor generation.[31, 32]

To enhance the light harvesting efficiency in DSSCs, plasmonic core-shell nanoparticles (PCSNPs) have recently been incorporated into the active layer of DSSCs.[16-22] While the metallic core concentrates light, the shell is ought to provide optimal spacing, chemical stability, good electrical conductivity as well as good photovoltaic properties. Therefore, the morphology of the core as well as the material and thickness of the shell need be carefully designed and controlled in order to obtain optimal enhancement. Typically, the enhancement effect of metal@dielectric core-shell nanoparticles is attributed to the enhanced excitation rate of dye sensitizers due to LSPR. When using semiconducting shell, such as $TiO_2$, two additional mechanisms may be involved in the efficiency enhancement. First, the hot electrons generated inside the resonant plasmonic cores can contribute directly to the photocurrent.[33, 34] Second, the metal cores can undergo charge equilibrium with the surrounding semiconductor and modify the Fermi level of $TiO_2$, resulting in an improved cell potential.[20] In addition, using $TiO_2$ as shell material may also benefit from the compatible affinity to dye sensitizers[35] and the reduced internal impedance of the cell.[19, 36, 37] At nanometer scale, varying the thickness of the very thin $TiO_2$ shell can, however, significantly change its semiconducting and optical properties, thereby altering the plasmonic enhancement effect in DSSCs. To obtain maximum enhancement, knowledge of influence of particle morphology on the enhancement effect as well as a reliable method for deterministic synthesis of designed PCSNPs are required.

In this work, we developed a new synthesis route for Au@$TiO_2$ PCSNPs to achieve deterministic and precise control of core morphology and shell thickness. With these well-defined PCSNPs, we systematically studied the influence of the thickness of $TiO_2$ shell on the plasmonic enhancement effect in DSSCs. We found that the enhancement mechanism varies with $TiO_2$ shell thickness. PCSNPs with thinner $TiO_2$ shell enhance short-circuit current due to plasmonic effect, whereas particles with thicker shell were found to enhance the open-circuit voltage due to semiconducting character of the shell. We show experimental evidence and discuss possible reasons for the observed difference. Our study gains insight into the plasmonic enhancement effect in DSSCs. The new synthesis method for well-controlled Au@$TiO_2$ PCSNPs may also find applications in solar-to-chemical energy conversion[40-42] and controlled nanoscale light-matter interaction.[43]

Octahedral gold nanoparticles were prepared following our previously reported method.[44] In short, $HAuCl_4 \cdot 3H_2O$ (0.01 mol/L, 5 mL, Alfa Aesar), Cetyltrime-thylammonium bromide (CTAB, 1.1 g, 98%, Alfa Aesar) and trisodiumcitrate (0.1 mol/L, 1.5 mL, 99.9%, Aldrich) were mixed with deionized water (Ultrapure Milli-Q water, R > 18 MΩ, 193.5 mL). The mixture solution was sealed and was heated to about but below boiling point for 12 hours.[44] The residual CTAB was removed by cleaning the sample several times with deionized water and centrifugation (6000 rpm, 30 minutes). The resulted citrate-capped gold nanoparticles were stored in deionized water for further $TiO_2$ shell synthesis. The number density of the as-prepared octahedral cores is around $1.37 \times 10^{14}$ particles/L calculated from the average particle size and the gold concentration obtained with ICP/MS (7500CE, Agilent).[39]

To synthesize $Au@TiO_2$ core-shell nanoparticles with well-controlled shell thickness, L-arginine aqueous solution (0.02 mol/L, 10 mL, Aldrich) was added into the octahedral gold cores solution (number density = $1.37 \times 10^{14}$ particles/L, 20 mL). 12 mL of cyclohexane was then added into the mixture to form a biphasic system. At this stage, the citrate-capped gold nanoparticles were turned into arginine-capped ones, which were further converted into thio-capped cores by adding 12 μL of (3-Mercaptopropyl)triethoxysilane (MPTS, 95%, Aldrich) into the cyclohexane layer and stirring for 6 hours.[45] In order to remove residual MPTS, cyclohexane layer was removed after 6 hours stirring and the particles were cleaned with dehydrated alcohol (Absolute ethanol, 99.8%, Aldrich) and centrifugation several times. At this stage, the gold cores were well dispersed in dehydrated alcohol and covered with a thin layer of MPTS which serves as an adhesion layer for further $TiO_2$ shell growth. It is worth noting that replacing the water solvent by dehydrated alcohol is necessary since the precursor of $TiO_2$ shell, i.e. titanium isopropoxide (TTIP), interacts instantly with water and forms $TiO_2$ nanoparticles instead of $TiO_2$ shell on gold cores in aqueous solution. To grow TiO$_2$ shell, the alcohol solution of gold cores was added 1μL of TTIP (97%, Aldrich) and allowed for growing under mild stirring for a controlled period of time ranging from 1.5 hours to 24 hours. The reaction was then terminated by rinsing the particles with dehydrated alcohol and centrifugation in order to remove residual TTIP. Scheme 1 summarizes the synthesis of the Au@TiO$_2$ core-shell nanoparticles. Since the growth of TiO$_2$ shell is very sensitive to water, it is important to maintain the humidity such that the TiO$_2$ shell grows in a steady and reasonably slow rate. This was done by controlling the temperature and humidity at 25 °C and 50%, respectively.

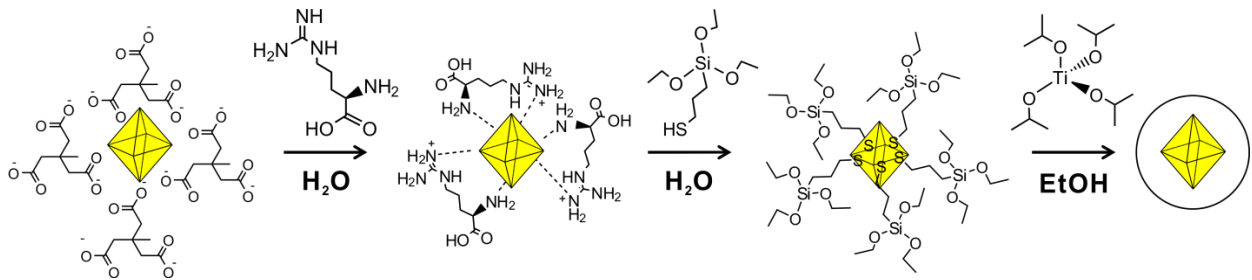

**Scheme 1.** Schematic diagram for the synthesis of Au@TiO$_2$ core-shell nanoparticles with well-controlled core morphology and shell thickness.

Figure 1a shows the average thickness of Au@TiO$_2$ nanoparticles as a function of reaction time. TiO$_2$ shell thickness first increases linearly with the reaction time at a slope of about 2.5 nm per hour and reaches a constant value of 37 nm due to depletion of TTIP. Based on such a linear relation, we were able to precisely tune the shell thickness by controlling the reaction time within the dynamic range (t < 10 hours). The shell thickness of the as-synthesized PCSNPs has a relatively narrow distributions (Figure 1S, Supporting Information).

Figure 1b shows the extinction spectra of nanoparticles measured with commercial spectrometer (V-570, Jasco). The extinction spectrum of gold cores clearly shows a plasmonic resonance peak around 545 nm. Upon addition of TiO$_2$ shell, the resonance shifts accordingly to longer wavelength (560-570 nm, indicated with an arrow) due to the higher index of the shell. Since the shell is thin and the field extends into the surrounding water, the observed shift is minute. To verify the experimentally observed resonance spectra, we performed simulations using finite-difference time-domain (FDTD, Lumerical Solutions, Canada) methods. In the simulations, average size of the particles obtained from field-emission scanning electron microscope (FE-SEM, JSM-7000F, JEOL) and field-emission transmission electron microscope (FE-TEM, JEM-2100, JEOL) images are used in order to mimic the real nanostructures. For bare gold octahedral cores, the simulated resonance position is in good agreement with the experimental data. However, for all the Au@TiO$_2$ core-shell nanoparticles, if the index of bulk TiO$_2$ material (n = 2.5) is assumed for the shell, the calculated resonance position exhibits significantly greater red shifts compared to the experimental data (Figure S2, Supporting Information). The discrepancy indicates that the effective index of the shell material is much lower than bulk TiO$_2$. In order to match the experimental resonance positions, the effective index used for the shell in FDTD simulations had to be lowered down to a value between 1.6 and 1.7, increasing with the shell thickness. The much lower and thickness-dependent effective index stems from the fact that surface of the TiO$_2$ shell is in the form of Ti-OH whose index is significantly lower than crystalline TiO$_2$ crystals.[46, 47] When the shell is thin, surface with Ti-OH units is the major constituent of the shell. Based on the mixing rule that the effective index of a composite material is proportional to the portion ratio of the constituent in the mixture,[48, 49] the index of the thin shell deviates from the value of bulk crystalline TiO$_2$. As a result, assuming a refractive index of bulk TiO$_2$ material for the shell in the simulation leads to significant deviation from the experimental data. In fact, such a discrepancy can also be seen in previously reported data.[41] Increasing the thickness of the shell, the contribution from the surface Ti-OH becomes less and less significant and the effective index of the shell gradually increases. Such a dependence of the shell constituent on its thickness has a major influence on the plasmonic enhancement effect in DSSCs, as will be discussed later. By FDTD simulations, we found that the effective index of the shell varies from 1.61 for 5-nm shell to 1.69 for 37-nm shell, as shown in the inset of Figure 1b. In fact, the increasing absorption with respect to the shell thickness seen in the short wavelength regime (< 500 nm) in Figure 1b also provides a direct evidence of the increasing portion of $TiO_2$ in thicker shell. As the thickness of the shell increases to a value larger than 40 nm, the resonance position slowly approaches 767 nm, which is the resonance of bare gold cores surrounded by bulk $TiO_2$ material, as in the active layer of a DSSC. As we will see later, it is such greatly red-shifted plasmonic resonance at the IR regime that shows pronounced enhancement effect in DSSCs.

Representative SEM images of the cores and FE-TEM images of various Au@$TiO_2$ core-shell nanoparticles are shown in Figure 1c. It can be clearly seen that the cores have well-defined octahedral shape and relatively narrow distribution of the side length (23.2 ± 1.1 nm). For the Au@$TiO_2$ core-shell nanoparticles, TEM images show that the gold cores are well covered by the shell with a time-dependent thickness. Since the Au@$TiO_2$ core-shell nanoparticles will eventually be doped into the $TiO_2$ paste for the active layer of DSSCs, we examined the morphology of the nanoparticles after being sintered at 500 ºC for one hour, as required in the fabrication of photoanodes. As shown in the bottom panel of Figure 1c, bare gold cores melt to form larger grains with irregular shape after sintering. On the contrary, the Au@$TiO_2$ core-shell nanoparticles maintain their morphology and the gold cores are still in octahedral shape (bottom panel of Figure 1c). Even though no crystalline structure was observed in the shell by high resolution TEM, the X-ray diffraction analysis (XRD-6000, Shimadzu) confirms that main constituent of the shell is TiO$_2$ based material (Figure S3, Supporting Information).

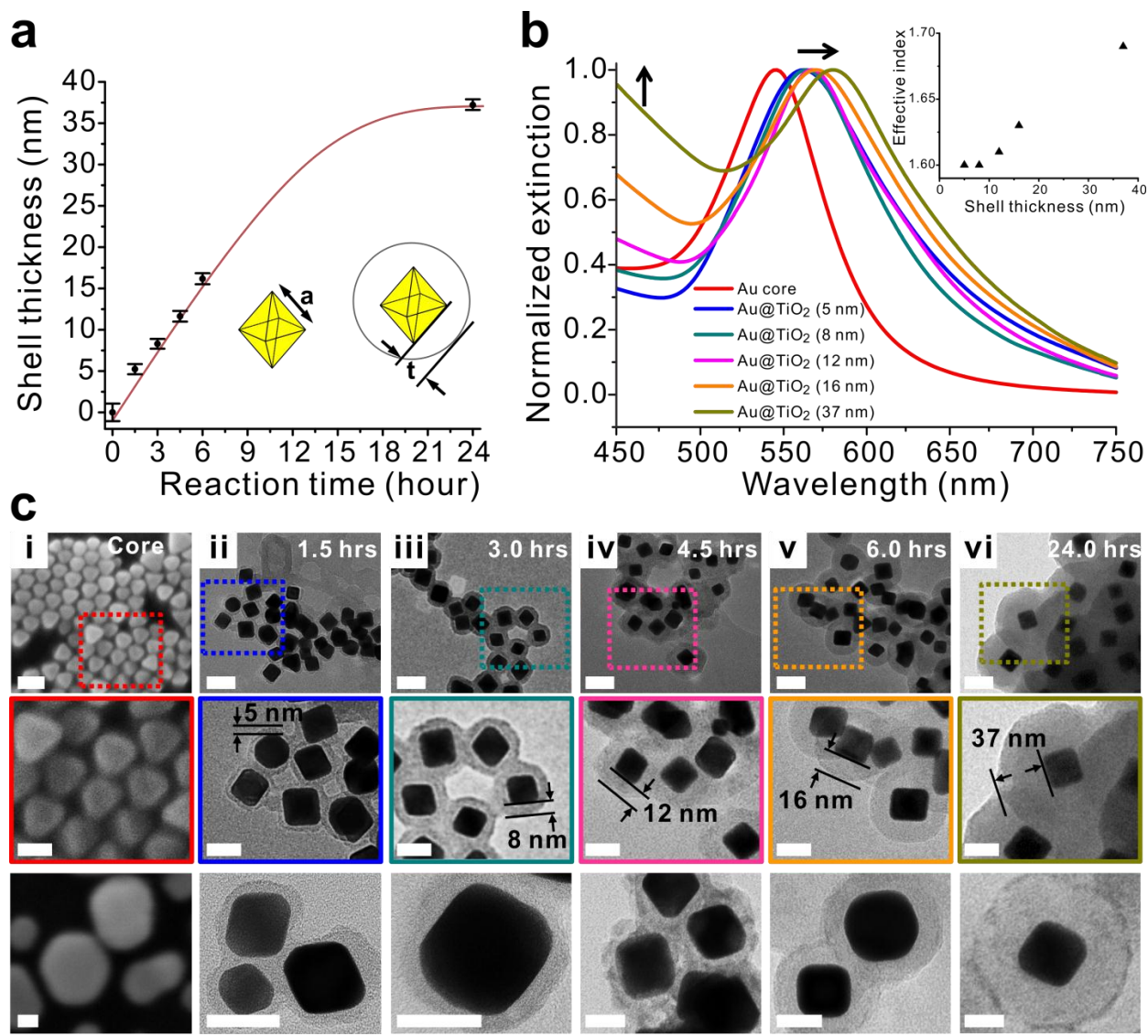

**Figure 1**. (a) Shell thickness, *t*, as a function of reaction time. At zero reaction time, the data point represents the distribution of the side length of octahedral cores, *a*, with an average value of 23.2 nm. Solid line is a guide for the eye. Inset illustrates how the shell thickness, *t*, and side length of the cores, *a*, are measured. Error bars are the standard deviations obtained from measuring 40 cores and 20 Au@TiO$_2$ nanoparticles using SEM and TEM images. (b) Extinction spectra of octahedral gold cores (red) and Au@TiO$_2$ nanoparticles with shell thickness of 5, 8, 12, 16, and 37 nm (blue, cyan, magenta, orange and dark yellow, respectively). The resonance position shifts to longer wavelength as the shell thickness increases. Inset shows the simulated effective index of the shell as a function of the thickness. (c) SEM images of the octahedral cores (i) and TEM images of the Au@TiO$_2$ core-shell nanoparticles with increasing reaction time and shell thickness (ii-vi). Middle panels show the zoomed-in images of the areas marked with dotted

squares in the corresponding top panels. Bottom panels show the images of corresponding nanoparticles after being sintered at 500 ℃ for 1 hour. Scale bars for top, middle and bottom panels are 50, 25 and 20 nm, respectively.

Being able to precisely control the thickness of $TiO_2$ shell, we then incorporated our PCSNPs into the active layer of DSSC and systematically studied their plasmonic enhancement effect on the cell performance. The assembly of DSSC devices used in this work followed the procedure reported by Chen et al.[50] except that the scattering layer was skipped and the active layer of the photoanode was modified by doping small amount of Au@$TiO_2$ PCSNPs into commercially available $TiO_2$ paste. The doped pastes were then used for the active layer of photoanode in DSSC devices (see Supporting information for details of DSSC assembly). The amount of doped Au@$TiO_2$ particles was carefully optimized to achieve maximum power conversion efficiency (PCE). The PCE is calculated using $\eta = \frac{J_{sc}V_{oc}FF}{I_s}$, where $J_{sc}$ is the measured short-circuit current, $V_{oc}$ is the open-circuit voltage, FF is the fill factor and $I_s$ = 100 mW/cm$^2$ is the irradiance of the light source in the measurement, in equivalent to one sun at air mass (AM 1.5) at the surface of the cells (Xenon lamp, Oriel 69907, Newport). To obtain the current-voltage characteristics of the cell, external bias voltage was applied to the DSSCs and the photocurrent was recorded with a digital source meter (Keithley model 2400, Keithley, U.S.A.). For each kind of PCSNPs, we fabricated ten replicates of DSSC devices in order to obtain the cell performance with statistical significance. Figure 2a shows the PCE of the solar cells as a function of doping level of Au@$TiO_2$ nanoparticles. It can be clearly seen that for all kinds of PCSNP dopants, the PCE first increases with the doping level and gradually decreases after reaching a maximum value. Carefully examining the curves of PCE with that of $J_{sc}$ (Figure 2b) and $V_{oc}$ (Figure 2c), two trends can be observed. First, for PCSNPs with thinner shell (thickness = 5, 8 and 12 nm), the trend of PCE follows that of $J_{sc}$, whereas the $V_{oc}$ is rather constant or irrelevant, indicating that the enhanced PCE is mainly due to increasing cell current. On the contrary, for PCSNPs with a 16-nm shell (orange traces in Figure 2), the improved PCE at lower doping level is apparently due to the increased $V_{oc}$, while $J_{sc}$ starts playing a role at middle doping level. Such observation reveals that the mechanism responsible for the enhancement effect of doping Au@TiO$_2$ PCSNPs depends on the thickness of the semiconducting shell. When using PCSNPs with thicker shell (t = 16 nm), the sensitizer molecules are kept farther from the cores and thus experience only marginal plasmonic field enhancement effect. As a result, $J_{sc}$ values for the cases of 16-nm shell are relatively lower (Figure 2b). Although thicker shell leads to lower current enhancement, the portion of crystalline TiO$_2$ in thicker shell is large enough to undergo charge equilibrium between metal cores and surrounding TiO$_2$ material. Consequently, an increased $V_{oc}$ and a slightly improved fill factor (Figure S4, Supporting Information) were obtained due to a modified Fermi level of the semiconductor.[20] The difference in the enhancement mechanism with respect to the shell thickness can be more clearly seen in the current-to-voltage (J-V) characteristics of the cell, as shown in Figure 2d. We compared the J-V characteristics of plasmonic DSSCs at optimal doping levels that give the best efficiency in Figure 2a. It is obvious that the $J_{sc}$ decreases as the shell thickness increases, whereas the $V_{oc}$ increases gradually with increasing thickness, as indicated by the black arrows in Figure 2d. This trend is in good accordance with the comparison made between Au@TiO$_2$ and Au@SiO$_2$ PCSNPs.[20] Considering the shell thickness-dependent constituent of the shell material, it is then clear that thinner shell prepared with our method shows no semiconducting character and serves as a dielectric layer as SiO$_2$.

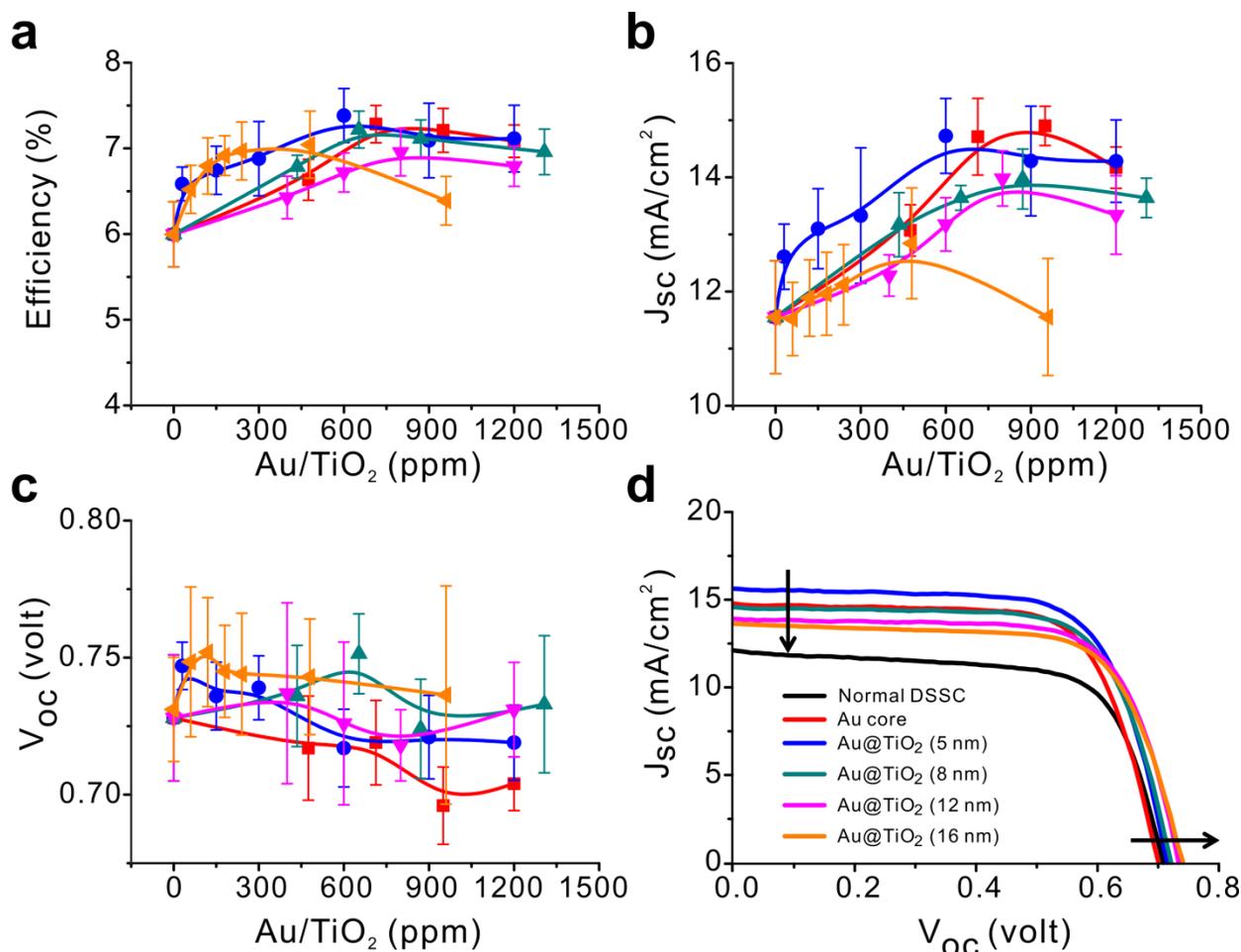

**Figure 2**. The influence of doping level of octahedral gold nanoparticles (red squares) and PCSNPs with 5-nm (blue dots), 8-nm (cyan up triangles), 12-nm (magenta down triangles) and 16-nm (orange left triangles) shell on (a) overall power conversion efficiency, (b) short-circuit current ($J_{sc}$), (c) open-circuit voltage ($V_{oc}$) and (d) current-voltage (J-V) characteristics of the DSSCs. In (a-c), the data points at zero doping level are the performances of normal DSSCs without doping plasmonic nanoparticles. Error bars are the standard deviation of 10 replicates for each device. In (d), only the J-V curves of plasmonic DSSCs with optimal doping levels that give the best efficiency in Figure 2a are plotted.

Table 1 summarizes the size and the resonance wavelength of the nanoparticles, as well as the best doping level and the resulted averaged performance of plasmonic DSSCs. Doping PCSNPs with 5-nm shell at a level of 600 ppm (gold-to-titania ratio determined by ICP/MS, 7500CE, Agilent) was found to give the best enhancement of PCE from 6.00% (normal DSSCs) to 7.38% (plasmonic DSSCs), corresponding to an efficiency improvement of 23%.

**Table 1.** Particle size, resonance wavelength, best doping level of PCSNPs, averaged performance of DCCSs and the corresponding efficiency enhancement compared to normal DSSCs.

| | thickness (nm) | resonance λ in $H_2O$ (nm) | doping (ppm) | $J_{sc}$ (mA/cm$^2$) | $V_{oc}$ (volt) | FF (%) | PCE (%) | enhancement of PCE (%) |
|---|---|---|---|---|---|---|---|---|
| normal DSSC | - | - | - | 11.55 ± 0.99 | 0.73 ± 0.02 | 0.71 ± 0.02 | 6.00 ± 0.38 | - |
| Au cores | 23.2 ± 1.1 (side length) | 545 | 713 | 11.47 ± 0.67 | 0.72 ± 0.02 | 0.70 ± 0.01 | 7.28 ± 0.22 | 21 |
| Au@TiO$_2$ (5 nm) | 5.2 ± 0.62 | 562 | 600 | 14.73 ± 0.65 | 0.72 ± 0.01 | 0.70 ± 0.01 | 7.38 ± 0.31 | 23 |
| Au@TiO$_2$ (8 nm) | 8.3 ± 0.59 | 564 | 653 | 13.64 ± 0.22 | 0.75 ± 0.02 | 0.70 ± 0.01 | 7.22 ± 0.22 | 20 |
| Au@TiO$_2$ (12 nm) | 11.6 ± 0.65 | 567 | 800 | 13.98 ± 0.48 | 0.72 ± 0.01 | 0.70 ± 0.01 | 6.95 ± 0.28 | 16 |
| Au@TiO$_2$ (16 nm) | 16.2 ± 0.67 | 570 | 480 | 12.84 ± 0.97 | 0.74 ± 0.02 | 0.71 ± 0.02 | 7.04 ± 0.39 | 17 |

In order to understand the plasmonic enhancement effect, we performed spectral analysis on the cell performance by means of incident photon-to-current conversion efficiency (IPCE) and carefully compared the spectral response of the cell with the surface plasmon resonance of the PCSNPs. Figure 3a shows the IPCE of the DSSCs with various PCSNPs dopants. The devices that give the best cell efficiency were used for IPCE analysis. It can be seen that the enhancement of the IPCE is universal over the whole extinction spectrum of the cell. The enhancement decreases, however, with increasing shell thickness. Since the near field responsible for the enhanced excitation of dyes is strongest at plasmonic resonance, enhancement in the cell current is expected to show spectral features that follows the surface plasmon resonance. To verify this point and gain insight into the enhancement effect, we plot the relative enhancement ($\Delta$ IPCE/IPCE) as a function of illumination wavelength, where $\Delta$ IPCE is the difference between the best plasmonic DSSCs and normal DSSCs. As can be seen in Figure 3b, pronounced enhancement is observed at the spectral regime above 750 nm, which well matches the plasmon resonance of nanoparticles in $TiO_2$ environment, as shown in Figure 3c. Note that in comparison with the resonance of particles in aqueous solution (Figure 1b), PCSNPs in $TiO_2$ paste significantly red shifts to the infrared (IR) regime since localized surface plasmon resonance greatly depends on the dielectric function of the environment.[22, 25] Around plasmon resonance peak, increasing the shell thickness decreases the enhancement in current. This is explained by the fact that plasmonic near-field enhancement decays rapidly form the gold surface. Since thicker shell keeps N719 molecules farther from the surface of the gold core, plasmonic enhancement effect is weaker. The relative IPCE enhancement is in very good agreement with the numerical simulation on the near-field intensity enhancement, as shown in Figure 3c. Intuitively, shell prevents the dye molecules from contacting the metal cores and helps avoid possible quenching of dye. Therefore, compared to bare cores, PCSNPs with thin shell should exhibit highest current enhancement. However, our results show that doping bare gold cores exhibits comparably large enhancement effect as PCSNPs with 5-nm shell, as can be seen in Figure 3a and 3b. The fact that dye molecules are in direct contact with the metal surface seems not playing important role and the quenching of dye molecules by the metal surface is insignificant in DSSCs. A possible reason for this can be the very efficient electron injection into the conduction band of the semiconductor, which is significantly faster than the quenching channel introduced by the metal surface. In order to understand the effect of the shell on the dye quenching, we performed emission lifetime measurement on N719 dye sensitizer in ethanol solution and in an isolated active layer on a cover glass (Figure S5, Supporting Information). We found that the originally long emission lifetime of N719 dyes in ethanol solution is greatly reduced to sub-nanosecond regime with the presence of $TiO_2$ nanoparticles. Further reduction of the emission lifetime was observed when small amount of plasmonic nanoparticles were doped into the $TiO_2$ paste. Compared to dye molecules in contact with the metal cores, N719 molecules neighboring PCSNPs with $TiO_2$ shell apparently exhibit longer lifetime, although the dependence of lifetime on the thickness is not obvious. This observation reveals that the metal surface of the cores does quench the excited dye molecules and lead to shorter emission lifetime. However, in a shorted DSSC, such quenching channels introduced by the metal surface are not fast enough to compete with the very efficient electron injection from the ligand of dye into the conduction band of $TiO_2$.[51] Therefore, no pronounced effect of the shell thickness can be seen from the performance of plasmonic DSSCs.

In addition to the enhancement around plasmonic resonance in IR spectral range, wavelength-independent enhancement in the visible (VIS) to ultraviolet (UV) spectral regime was also observed (inset of Figure 3b). In UV/VIS regime, the enhancement shows no significant wavelength-dependent features but decreases as the shell thickness increases. The shell thickness dependence indicates that the effect indeed stems from the doping of PCSNPs. The lack of wavelength dependence implies, however, that the enhancement effect is not directly due to enhanced light harvesting of dye molecules at this spectral regime, which would otherwise manifest itself as a broad peak at the plasmon resonance. Indeed, when using plasmonic nanoparticles, the enhancement of $J_{sc}$ in the non-resonance spectral regimes are commonly observed.[17, 19, 21] Here we tentatively attribute such a wavelength-independent enhancement to plasmonic heating effect.[29, 31, 32] It is well known that upon resonance, plasmonic nanoparticles can efficiently convert the electromagnetic energy to heat and raise the local temperature. This effect has been used to release drug and kill cancer cells locally.[31] Recently, it has been demonstrated that plasmonic nanoparticles can convert solar energy to localized heat and efficiently generate water vapor steam.[32] Such plasmonic heating effect should also occur in plasmonic DSSC devices and enhance the cell performance by promoting the electrochemical reaction in the cell or increasing the mobility of electrolyte,[52] leading to an enhancement in $J_{sc}$ without spectral dependence. For DSSCs doped with PCSNPs with thicker shell, such effect is less pronounced since the heat produced in the metal cores is more isolated. As a result, the enhancement in the IPCE and $J_{sc}$ decreases as the shell thickness increases. Other possible reason for universal enhancement, such as increased dye loading or enhanced conductivity, should be relatively minor since the doping level is small and the surface affinity of the Au@TiO$_2$ is similar to the TiO$_2$ particles in the paste.

Finally, we would like to point out that the plasmonic enhancement around 750 nm in Figure 3b is pronounced because the plasmon resonance is at the edge of dye absorption spectrum. While it is beneficial to observed plasmonic enhancement effect at low absorption spectral regime,[24] it is also our intention to use IR plasmonic nanoantennas in solar cells to improve the absorption of light at this weakly absorbed spectral region, as in the trend of developing red absorbing sensitizers.

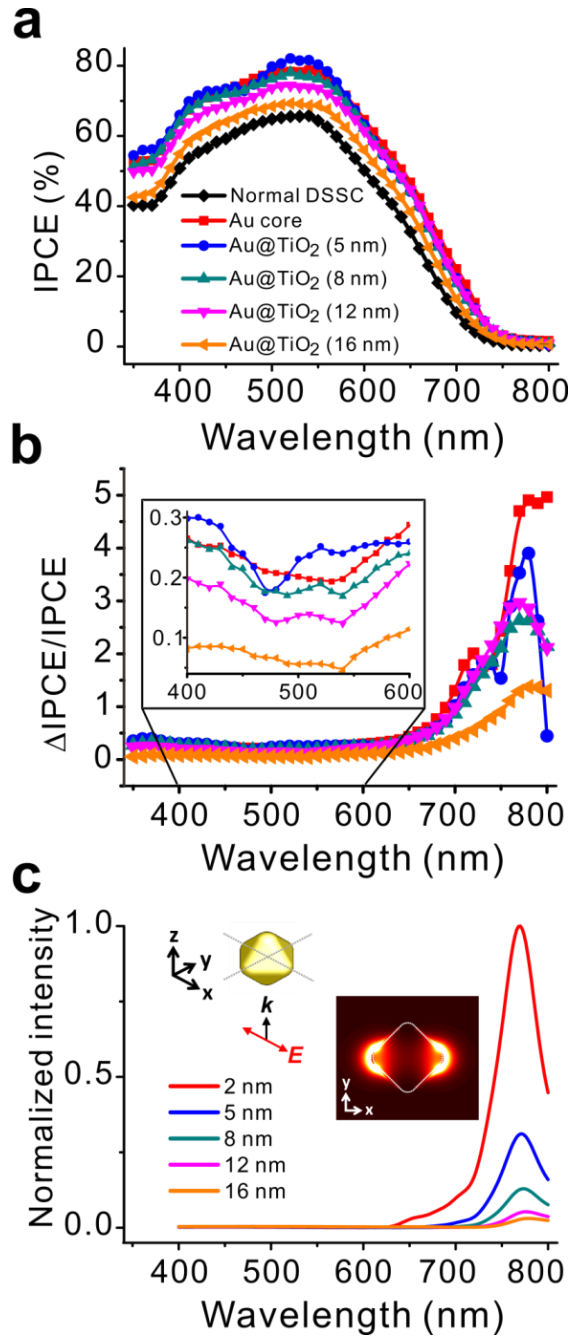

**Figure 3**. (a) IPCE and (b) relative enhancement in IPCE (△IPCE/IPCE) of DSSCs without doping (black), with doing bare gold cores (red), PCSNPs with shell thickness of 5 nm (blue), 8 nm (cyan), 12 nm (magenta) and 16 nm (orange). Inset shows the zoom-in of spectral range of 400 to 600 nm. For doped DSSCs, optimal doping levels shown in Figure 2a are used. (c) Simulated normalized near-field intensity enhancement at resonance wavelength of 766 nm, recorded at a position that is 2 nm (red), 5 nm (blue), 8 nm (cyan), 12 nm (magenta) and 16 nm (orange) from the surface of gold core. Inset shows the polarization of electric field of the incident excitation as well as the near-field enhancement distribution recorded at 766 nm.

In conclusion, we have systematically studied the influence of the shell thickness on the enhancement effect in plasmonic DSSCs using Au@TiO$_2$ core-shell nanoparticles. We have developed a new route for the synthesis of well-controlled Au@TiO$_2$ particles. We found that the enhancing mechanism varies with the shell thickness. Doping PCSNPs with thinner TiO$_2$ shell enhances the excitation rate of dye sensitizers and increases the cell temperature, leading to enhancement in the short-circuit current. On the other hand, doping PCSNPs with thicker shell allows for charge equilibrium between semiconductor and the metal cores, resulting in an increased open-circuit voltage. Maximum plasmonic enhancement effect was observed in the IR regime. Wavelength-independent enhancement was observed in the visible to ultraviolet spectral regime, which we attribute it to plasmonic heating effect. Emission lifetime measurement for N719 dyes confirms that TiO$_2$ shell play role in preventing the dye molecules from being quenched by the metal surface. The effect of quenching by the metal core is, however, insignificant in DSSCs since the electron injection to semiconductor is much more efficient. In order to obtain best enhancement, the absorption band of dyes and the resonance of the plasmonic nanoparticles should be tuned to overlap and preferably in the red to IR spectral regime such that this portion of solar spectrum can be efficiently absorbed. Our work gains insight into the plasmonic enhancement effect in DSSCs and provides new method for the synthesis of Au@TiO$_2$ PCSNPs with well-controlled morphology which may find applications in solar-to-chemical energy conversion and controlled nanoscale light-matter interaction.

ASSOCIATED CONTENT

**Supporting Information**. Assembly of DSSCs, particle size distribution, numerical simulations, XRD spectra, fill factor of DSSCs, N719 emission lifetime.

AUTHOR INFORMATION

**Corresponding Author**

*Email: jshuang@mx.nthu.edu.tw

**Notes**

The authors declare no competing financial interest.

ACKNOWLEDGMENT

The authors thank I.-C. Chen, J.-T. Lin and L.-L. Li for fruitful discussions and A.-T. Lee for the assistance in experiments. Supports from National Science Council of Taiwan under Contract Nos. NSC-99-2113-M-007-020-MY2 and NSC-101-2113-M-007-002-MY2 are gratefully acknowledged.

# Supporting Information

# Influence of morphology on the plasmonic enhancement effect of Au@TiO$_2$ core-shell nanoparticles in dye-sensitized solar cells

Wei-Liang Liu,[1] Fan-Cheng Lin,[1] Yu-Chen Yang,[2] Chen-Hsien Huang,[1] Shangjr Gwo,[2,3] Michael H. Huang[1] and Jer-Shing Huang[1,3,4*]

1. Department of Chemistry, National Tsing Hua University, Hsinchu 30013, Taiwan

2. Department of Physics, National Tsing Hua University, Hsinchu 30013, Taiwan

3. Center for Nanotechnology, Materials Sciences, and Microsystems, National Tsing Hua University, Hsinchu 30013, Taiwan

4. Frontier Research Center on Fundamental and Applied Sciences of Matters, National Tsing Hua University, Hsinchu 30013, Taiwan

* jshaung@mx.nthu.edu.tw

# Assembly of DSSCs

In this work, DSSC devices were prepared following the procedure described in [ref. S1]. The only difference is that the scattering layer was skipped and various plasmonic nanoparticles were doped into commercially available $TiO_2$ paste (Ti-Nanoxide T/SP, Solarnix). Briefly, $TiO_2$ paste was first mixed with various PCSNPs and then blade-coated [ref. S2] onto FTO glass substrates to form the active layer with an area of 4×4 mm$^2$. The photoanodes were then dried and baked in an oven with the following temperature program: (1) stage 1: raising temperature at a speed of 5 ℃/min until 80 ℃; (2) stage 2: drying the sample at 80 ℃ for 2 hours; (3) stage 3: raising temperature at a speed of 5 ℃/min until 500 ℃; (4) stage 4: baking the sample at 500 ℃ for 1 hour; (5) stage 5: cooling down the sample naturally in the oven until room temperature. The thickness of the as-prepared active layer is about 10 μm, as determined by a profilometer (Alpha-Step IQ, KLA-Tencor). The photoanodes were then dipped into anhydrous ethanol solution of N719 dyes (5×10$^{-4}$ M) for 18 hours to allow for maximum absorption of dyes. The photoanodes were then rinsed with anhydrous ethanol to remove residual dye solution. Platinized FTO counter electrodes were prepared by dropping 150 μL $H_2PtCl_6$ ethanol solution onto FTO glasses. After drying under room temperature for 30 minutes, the $H_2PtCl_6$ on counter electrodes was thermopyrolyzed using the following temperature program: (1) stage 1: raising temperature at a speed of 5 ℃/min until 80 ℃; (2) stage 2: drying the sample at 80 ℃ for 30 minutes; (3) stage 3: raising temperature at a speed of 5 ℃/min until 400 ℃; (4) stage 4: baking the sample at 400 ℃ for 15 minutes; (5) stage 5: cooling down the sample naturally in the oven until room temperature. Polyimide tape (60 μm thick, Taimide Tech. Inc.) as a spacer layer was then used to define a 4×4 mm$^2$ aperture on the counter electrode in order to match the size of the working area of photoanode. Electrolyte solution containing 0.5 M LiI, 0.05 M $I_2$ and 0.5 M 4-tert-butylpyridine in anhydrous acetonitrile was then injected into the aperture of the counter electrodes. The photoanodes and counter electrodes were then adhered and clamped with clips to form home-made DSSC devices used in this work.

# Particle size distribution

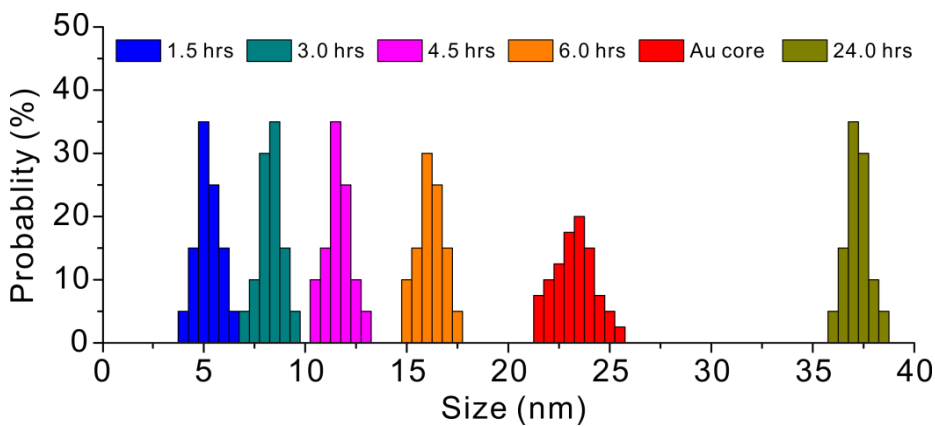

**Figure S1.** Statistics on the size distribution of PCSNPs. For octahedral gold cores (black), side lengths of forty particles are measured. For Au@TiO$_2$ nanoparticles, the thickness of the shell for twenty particles grown with 1.5 (blue), 3.0 (cyan), 4.5 (magenta), 6.0 (orange) and 24.0 hours (dark yellow) are measured from TEM images.

# Numerical simulations

The resonance position of the synthesized PCSNPs was simulated by finite-difference time-domain (FDTD) methods using average dimensions obtained from TEM images. To simulate the extinction spectra of PCSNPs dispersed in water, refractive index of the surrounding medium was set to be a constant of 1.33 within the frequency window of interest. The dielectric functions of gold and bulk TiO$_2$ material are obtained by using a multi-coefficient model [ref. S3] to fit the experimental data for gold [ref. S4] and TiO$_2$. [ref. S5] All boundaries of the simulation box were set to be 1000 nm away from the outer surface of the shell in order to avoid spurious absorption of the antenna near fields. A uniform mesh volume with discretization of 0.6×0.6×0.6 nm$^3$ covering the whole core-shell nanoparticles was used to describe the curved surface and to obtain converging simulation results within available computational ability. Corners of the octahedral gold core are rounded using cylinders with 6 nm diameter. A 2.7-fs pulsed plane wave source centered at 600 nm was set as the source for both far field extinction spectrum and near-field intensity simulations. The polarization of the source was aligned to the diagonal of the octahedron as shown in the inset of Fig. 3c in the main text. Directly using the index of bulk TiO$_2$ material for the thin shell leads to significant deviation from the experimental results, as shown by the blue stars in Fig. S2a. In order to find the effective index of the TiO$_2$ shell, the index of shell in FDTD simulations was varied between 1.6 and 1.7 until the simulated resonance positions (red dots, Fig. S2a) match the corresponding experimental data (black squares, Fig. S2a). As the thickness of the shell increased to a value larger than 40 nm, the resonance slowly approaches 767 nm, which is the resonance of bare gold cores surrounded by bulk TiO$_2$ material, as in the active layer of a DSSC. Such resonance position is good agreement with the plasmonic enhancement seen in the relative IPCE plot, as shown in Fig. 3b in the main text. One dimensional near-field intensity enhancement at the fundamental resonance (766 nm) of nanoparticle in TiO$_2$ environment is shown in Fig. S2b. The near-field intensity enhancement decreases rapidly from the gold/TiO$_2$ interface, in accordance with the observation seen in relative IPCE enhancement (Fig.3b in the main text).

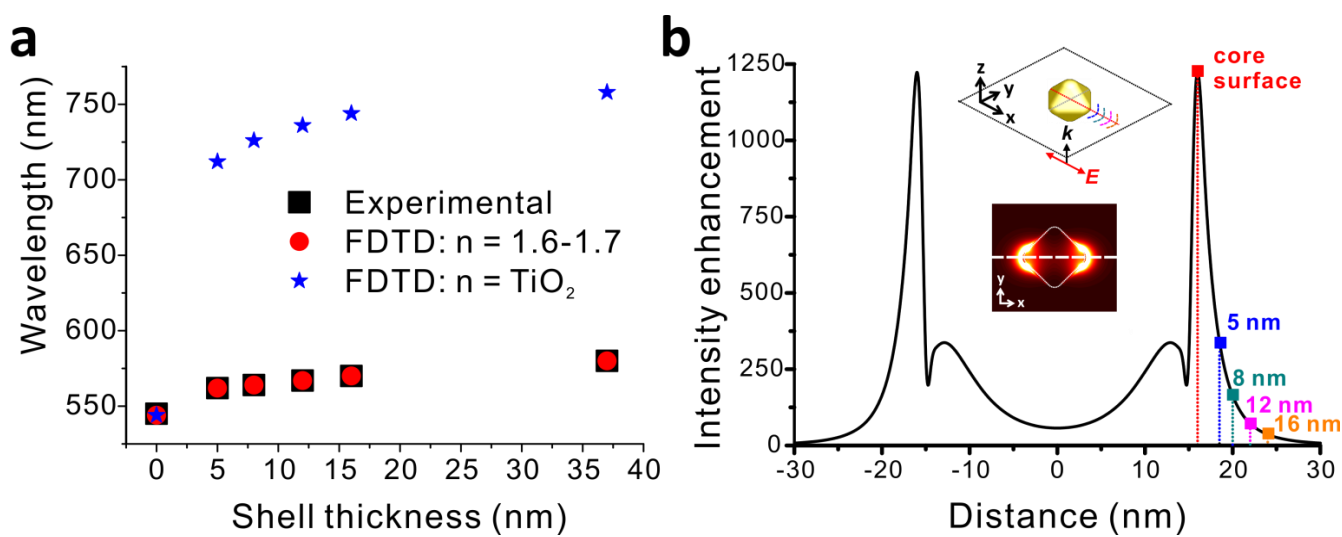

**Figure S2.** (a) Resonance wavelength of PCSNPs in water obtained from extinction experiment (black squares) and with FDTD simulations assuming an effective refractive index of bulk TiO$_2$ material (blue stars). For data points presented by red dots, effective index of the shell was tuned such that the simulated resonance overlaps with the experimental results. (b) Line-cut of near-field intensity enhancement of a nanoparticle in a TiO$_2$ environment. The insets illustrate the excitation polarization used in the simulation and the two-dimensional intensity enhancement distribution with a dashed line, along which the enhancement is plotted. The rapid decay of the near field is marked with colored solid squares and the distance from metal surface.

# XRD spectra

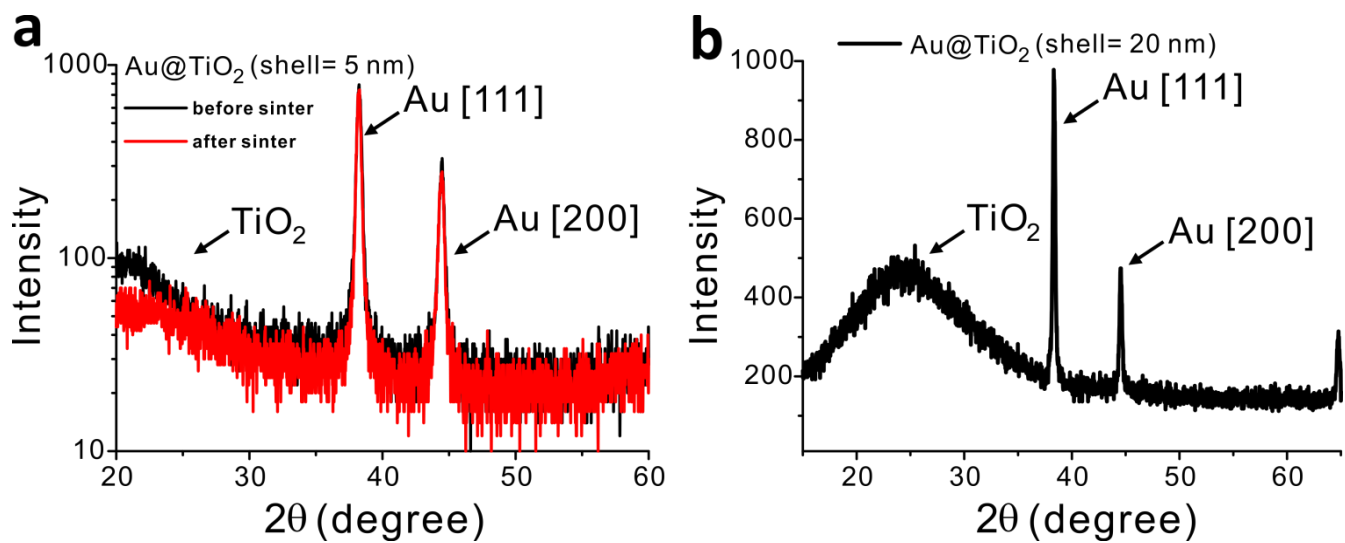

**Figure S3.** (a) XRD spectra (Shimadzu XRD-6000) of Au@TiO$_2$ PCSNPs (shell = 5 nm) before (black) and after (red) sintering under 500 ℃ for 1 hour. The signal below 30 degree corresponds to amorphous TiO$_2$ material. (b) XRD spectrum of Au@TiO$_2$ PCSNPs (shell = 20 nm) before sintering. The TiO$_2$ signal is more pronounced due to more TiO$_2$ material in the thicker shell.

# Fill factor of DSSCs

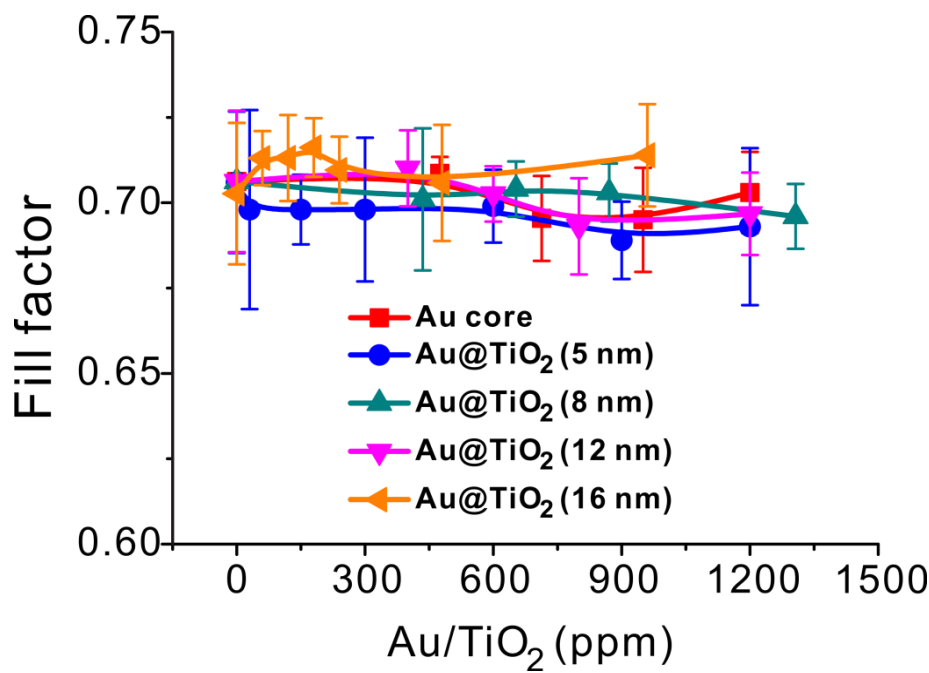

**Figure S4.** The influence of doping level of octahedral gold nanoparticles (red squares) and PCSNPs with 5-nm (blue dots), 8-nm (cyan up triangles), 12-nm (magenta down triangles) and 16-nm (orange left triangles) shell on the fill factor of the DSSC devices.

# N719 emission lifetime

Lifetime measurement was performed on N719 dyes in ethanol solution ($5\times10^{-4}$M), dried on cover glasses and absorbed in the active layer of solar cell photoanode. The instrumental setup for lifetime measurement is summarized in the Figure S6. Briefly, 10-picosecond pulses centered at 532 nm generated from a Nd:YVO$_4$ laser (High Q Laser, picoTRAIN) with repetition rate of 26.6 MHz and averaged power of 1 mW were focused onto the samples. Laser scattering was rejected by dichroic mirror (DCM13, Newport) and a long pass filter (FCL550, Thorlab). The emission from N719 dyes are aligned into an avalanche photodiode detector (PDM, Micro Photon Devices) with a time resolution of 50 ps and quantum efficiency of about 15% @ 780 nm, i.e. the emission wavelength of N719 dyes. The photon arrival time with respect to the excitation time was recorded and analyzed with a time-correlated single photon counting module (HydraHarp 400, PicoQuant). Lifetime fitting was performed using a complex multi-exponential fitting function (SymPhoTime, PicoQuant) that take into account the instrumental response function (IRF). The fitted lifetime for free N719 molecules in ethanol solution is 9.2 ns. As the dye solution was drop casted on the surface bare cover glass, the lifetime reduced to about 5 ns. For N719 dye molecules absorbed in a the active layer of normal DSSCs prepared with commercially available anatase TiO$_2$ paste, the lifetime further reduced to 602 ps. With gold core particles doped into the paste, the N719 lifetime dramatically reduced to less than 40 ps. For PCSNPs, the lifetime slightly recovered to about 60 ps. It worth noting that due to the fact that the response time of photon detector is about 50 ps, we are not able to accurately obtain the lifetime of the dye molecules in the vicinity of the cores or PCSNPs. Nevertheless, the longer lifetime for PCSNPs in comparison to that for bare cores is evident. More accurate lifetime measurement using transient absorption scheme are to be carried out.

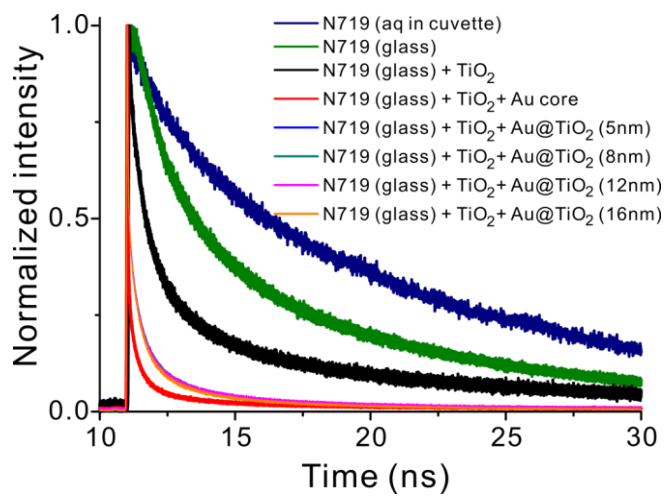

**Figure S5.** Emission lifetime of N719 sensitizer dye in aqueous solution (navy), dried on cover glass (green), dried on cover glass coated with sintered $TiO_2$ paste (black) and dried on cover glass coated with sintered $TiO_2$ paste mixed with octahedral gold cores (red), PCSNPs with shell of 5 nm (blue), 8 nm (cyan), 12 nm (magenta) and 16 nm (orange) thickness. For doped DSSCs, optimal doping levels shown in Figure 2a in the main text are used.

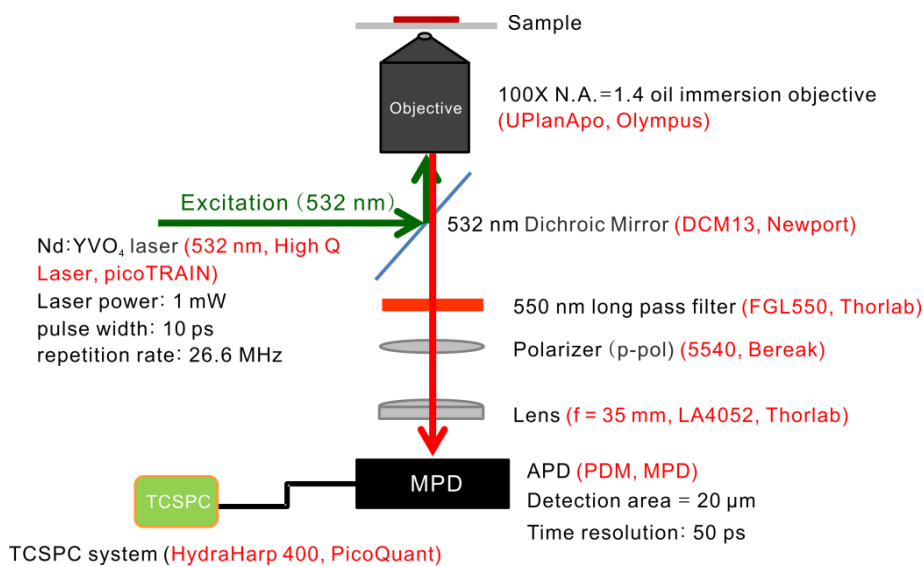

**Figure S6.** Schematic of the setup used for the emission lifetime measurement.

# Supporting References